\documentclass[12pt]{article}
\usepackage{graphicx}
\usepackage{wrapfig}
\usepackage{xspace}
\usepackage{amsmath,amssymb}
\usepackage{lineno}

\def\antibar#1{\ensuremath{#1\bar{#1}}}%

\newcommand*{\qqbar}{\antibar{q}}

\newcommand*{\nnbar}{\antibar{\nu}}

\newcommand*{\ttbar}{\antibar{t}}

\newcommand*{\bbbar}{\antibar{b}}
\newcommand*{\ee}{\ensuremath{e^{+} e^{-}}\xspace}

\newcommand*{\mumu}{\ensuremath{\mu^{+} \mu^{-}}\xspace}

\newcommand*{\Zboson}{\ensuremath{Z}\xspace}
\newcommand*{\Wplus}{\ensuremath{W^{+}}\xspace}
\newcommand*{\Wminus}{\ensuremath{W^{-}}\xspace}
\newcommand*{\Wboson}{\ensuremath{W}\xspace}

\let\pii=\pi
\renewcommand*{\pi}{\ensuremath{\pii}\xspace}

\newcommand*{\piplus}{\ensuremath{\pii^+}\xspace}

\let\etaa=\eta
\renewcommand*{\eta}{\ensuremath{\etaa}\xspace}

\newcommand*{\Kminus}{\ensuremath{K^-}\xspace}

\let\psii=\psi  
\renewcommand*{\psi}{\ensuremath{\psii}\xspace}

\newcommand*{\pT}{\ensuremath{p_{\text{T}}}\xspace}

\newcommand*{\ET}{\ensuremath{E_{\text{T}}}\xspace}

\newcommand*{\MET}{\ensuremath{E_{\mathrm{T}}^{\text{miss}}}\xspace}

\newcount\hrs\newcount\minu\newcount\temptime
\def\hm{\hrs=\time \divide\hrs by 60 \minu=\time\temptime=\hrs
\multiply\temptime by 60%
\advance\minu by -\temptime
\ifnum\minu<10 \let\zerofill=0\else \let\zerofill=\relax\fi
 \the\hrs:\zerofill\the\minu}

\newcommand*{\lapprox}{\ensuremath{\sim\kern-1em\raise 0.65ex\hbox{$<$}}\xspace}
\newcommand*{\rapprox}{\ensuremath{\sim\kern-1em\raise 0.65ex\hbox{$>$}}\xspace}

\newcommand*{\rts}{\ensuremath{\sqrt{s}}\xspace}
\newcommand*{\stat}{\mbox{$\;$(stat.)}\xspace}
\newcommand*{\syst}{\mbox{$\;$(syst.)}\xspace}
\newcommand*{\TeV}{\ifmmode {\mathrm{\ Te\kern -0.1em V}}\else
                   \textrm{Te\kern -0.1em V}\fi}%
\newcommand*{\GeV}{\ifmmode {\mathrm{\ Ge\kern -0.1em V}}\else
                   \textrm{Ge\kern -0.1em V}\fi}%
\newcommand*{\MeV}{\ifmmode {\mathrm{\ Me\kern -0.1em V}}\else
                   \textrm{Me\kern -0.1em V}\fi}%
\newcommand*{\keV}{\ifmmode {\mathrm{\ ke\kern -0.1em V}}\else
                   \textrm{ke\kern -0.1em V}\fi}%
\newcommand*{\eV}{\ifmmode  {\mathrm{\ e\kern -0.1em V}}\else
                   \textrm{e\kern -0.1em V}\fi}%

\newcommand*{\ifb}{\mbox{fb$^{-1}$}}
\newcommand*{\ipb}{\mbox{pb$^{-1}$}}

\newcommand\pubnumber{}
\newcommand\pubdate{\today}

\def\napoli{Department of Physics\\
Tokyo Institute of Technology, 152-8551 Tokyo, JAPAN}

\def\Title#1{\begin{center} {\Large #1 } \end{center}}
\def\Author#1{\begin{center}{ \sc #1} \end{center}}
\def\Address#1{\begin{center}{ \it #1} \end{center}}

\newcommand\pubblock{\rightline{\begin{tabular}{l} \pubnumber\\
         \pubdate  \end{tabular}}}
\newenvironment{Abstract}{\begin{quotation}  }{\end{quotation}}
\newenvironment{Presented}{\begin{quotation} \begin{center} 
             PRESENTED AT\end{center}\bigskip 
      \begin{center}\begin{large}}{\end{large}\end{center} \end{quotation}}

\begin{document}
\begin{titlepage}
\pubblock

\vfill
\Title{\Wboson, \Zboson and photon production at the LHC}
\vfill
\Author{ Masahiro Kuze}
\center{\small(on behalf of ATLAS and CMS Collaborations)}
\Address{\napoli}
\vfill
\begin{Abstract}
Recent results on \Wboson, \Zboson and photon production at the Large Hadron Collider
are presented.  Inclusive \Wboson and \Zboson/$\gamma^*$ production, their
production in association with jets and heavy flavors, and prompt photon, 
$\gamma\gamma$ and $\gamma$+jets production are discussed.
\end{Abstract}
\vfill
\begin{Presented}
XXXIV Physics in Collision Symposium \\
Bloomington, Indiana,  September 16--20, 2014
\end{Presented}
\vfill
\end{titlepage}
\def\thefootnote{\fnsymbol{footnote}}
\setcounter{footnote}{0}

\section{Introduction}
During the LHC Run 1 data-taking, both ATLAS and CMS experiments collected 
$pp$ collision data corresponding to approximately 5~\ifb~at \rts = 7~\TeV
~(up to 2011) and 20~\ifb~at \rts = 8~\TeV~(in 2012).
Among the various studies of Standard Model (SM) processes,
\Wboson, \Zboson and $\gamma$ productions are particularly interesting
since they involve electoweak probes in $pp$ interaction which is dominated
by the strong interaction.  They give clean experimental signatures with leptonic
decay modes (including a neutrino inferred by missing transverse energy, \MET)
and serve as benchmarks of SM validation at the highest energy.
The large masses of \Wboson and \Zboson assure the existence of a hard scale
that justifies the predictions based on perturbative QCD (pQCD).
The measurements span to extreme kinematics (transverse momentum
\pT up to $\approx$ 1~\TeV) or topologies (like number of jets) and
they allow tuning of tools for SM predictions, such as Next-to-Leading Order (NLO)
calculations and Monte Carlo (MC) event generators.
All these validations are crucial in the searches for beyond-SM physics,
the signal of which often involves \Wboson/\Zboson/$\gamma$ (and many
jets/\MET).
\section{Inclusive \Wboson, \Zboson/$\gamma^*$ production}
CMS made a measurement of inclusive \Wboson and \Zboson production at 8~\TeV
~using a special data set corresponding to 18.2~\ipb~with low pile-up (multiple $pp$ collisions occurring in the
same bunch crossing)~\cite{p7}.
The ratio of the inclusive cross sections is found to be
$R_{\Wboson/\Zboson} = 10.63 \pm 0.11 \stat \pm 0.25 \syst$, while the SM prediction is
$10.74 \pm 0.04$ using the FEWZ~\cite{FEWZ} NNLO (Next-to-NLO) calculation with the MSTW2008~\cite{MSTW} Parton Distribution Function (PDF).
The ratio of \Wplus and \Wminus cross sections is found to be
$R_{\Wplus/\Wminus} = 1.39 \pm 0.01 \stat \pm 0.02 \syst$ to be compared with the prediction
$1.41 \pm 0.01$.  Both measurements agree well with SM.

The LHCb experiment also measured \Wboson production using its detector that has an acceptance
in the forward region (pseudorapidity of the muon $2 < \eta(\mu) < 5$, where $\eta = \ln \tan (\theta/2)$ with 
$\theta$ being the polar angle with respect to the beam direction)~\cite{p8}.
This measurement nicely extends the ATLAS/CMS measurements in the central pseudorapidity region
and is sensitive to PDF predictions.

In the Drell-Yan process, \qqbar~annihilate to $\gamma^{*}/\Zboson$ (they have the same quantum numbers and thus interfere) and convert/decay to \ee or \mumu.  CMS measured the 8~\TeV~cross section in a wide mass range~(see Fig.~\ref{highDY}) and also the double-differential cross sections in mass and lepton-pair rapidity~\cite{p9-1}.
ATLAS has performed Drell-Yan measurements in the low mass region, also including the 2010 data set in which
the energy thresholds for leptons were as low as 6 and 9~\GeV.  In the lowest invariant-mass bin, the measurement
agrees better with the prediction from NNLO calculation than that from NLO (see Fig.~\ref{lowDY})~\cite{p9-2}.
        \begin{figure}[h]
			\vspace{-5mm}
		         	\begin{minipage}[t]{0.49\linewidth}
			\vspace{-7.3cm}
			\includegraphics[width=\linewidth,angle=270]{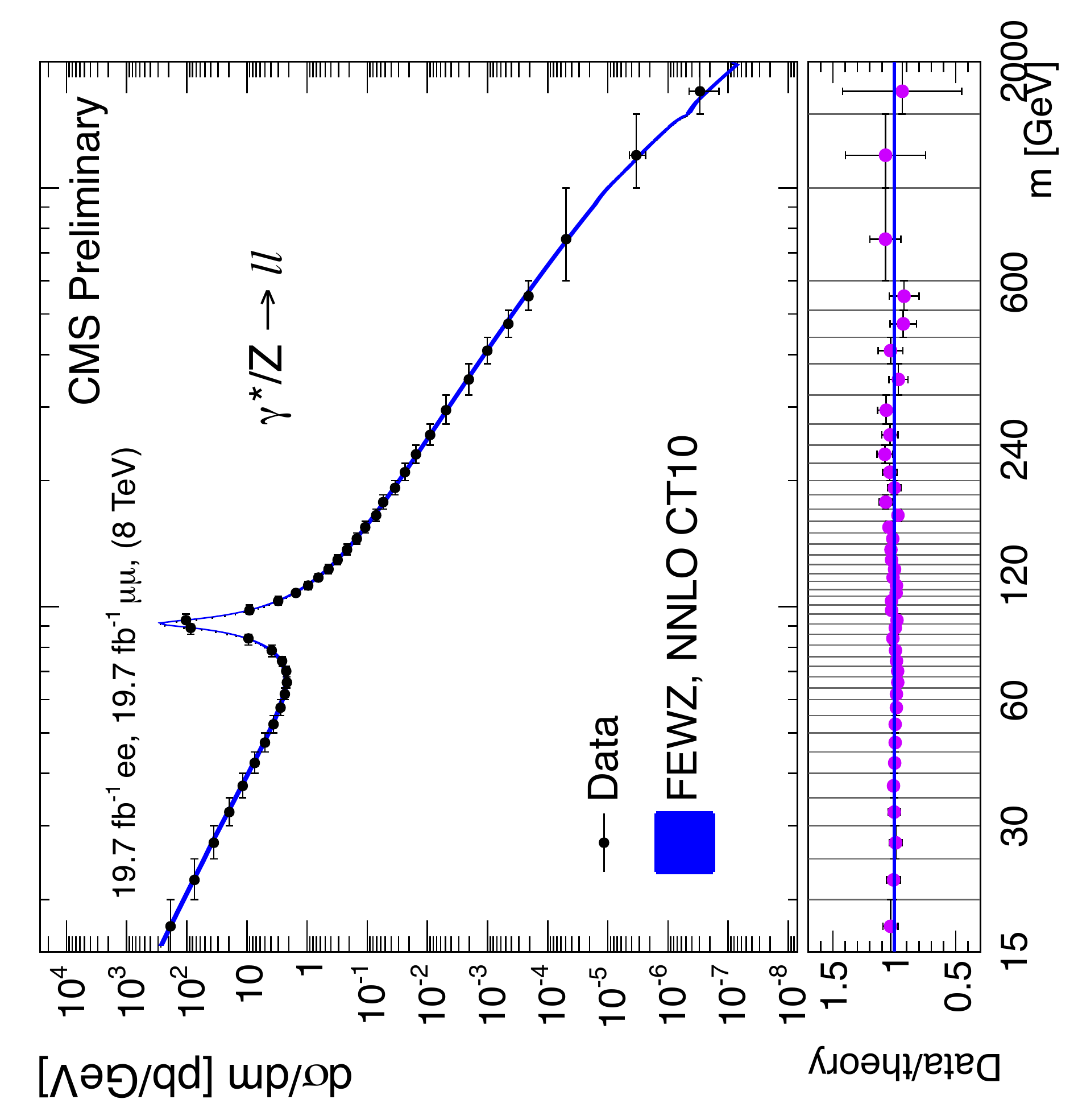}
			\caption{Drell-Yan differential cross section as a function of the di-lepton mass~\cite{p9-1}.}
			\label{highDY}
		\end{minipage}
		\hspace{0.01\linewidth}
		         	\begin{minipage}[t]{0.49\linewidth}
			\includegraphics[width=\linewidth]{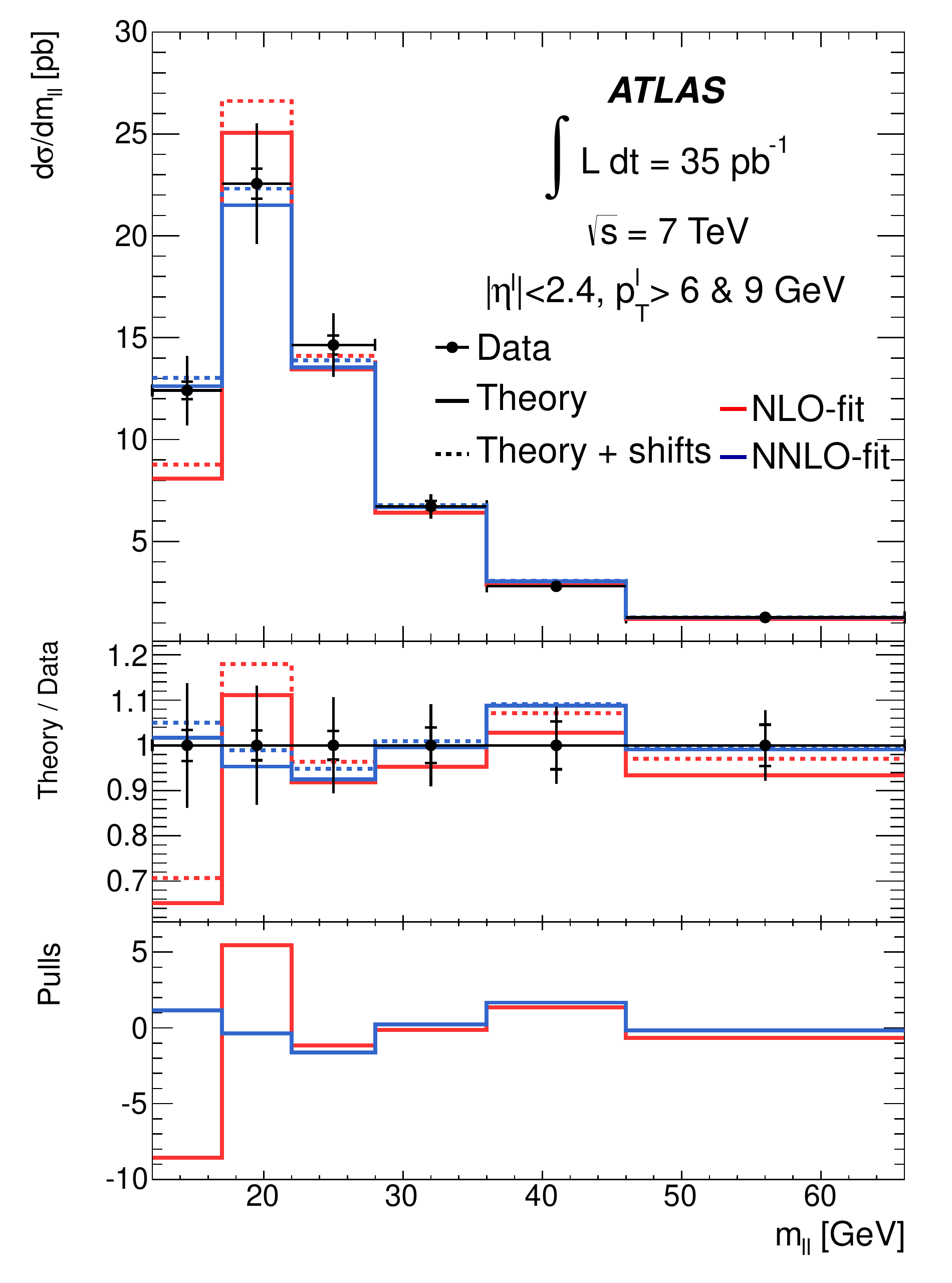}
			\caption{Low-mass Drell-Yan cross section compared to NLO and NNLO predictions~\cite{p9-2}.}
			\label{lowDY}
		\end{minipage}
         \end{figure}	
         
In the \Zboson production, the \pT and rapidity distributions were measured by CMS~\cite{p10-1}
using the 8~\TeV~data.
In the large \pT region of \Zboson~boson, the data show a softer distribution than the prediction.
ATLAS used the measurements in the low \pT region ($<$ 26~\GeV) to tune
the MC generators (PYTHIA8~\cite{PYTHIA8} and POWHEG~\cite{POWHEG}+PYTHIA8) with respect to 
the parton shower parameters~\cite{p10-2}.
The tuned MC prediction agrees with the measurement up to 50~\GeV~better than 2\%.

Although the leptonic decays are usually used to reconstruct \Zboson bosons, 
one can also see them in hadronic decay mode, despite the large QCD multijet background.
ATLAS used $b$-tag to suppress background and measured the $\Zboson \to \bbbar$ signal
in the region of di-jet \pT $>$ 200~\GeV~\cite{p11-1}.
ATLAS also analyzed the jet substructure and obtained the signal of \Wboson/\Zboson decay to \qqbar~in the distribution of the jet mass (see Fig.~\ref{jetmass})~\cite{p11-2}.
Boosted \Wboson/\Zboson will be important in searches for heavy particles.

The $\Zboson \to 4\ell$ peak was already seen in the Higgs observation papers~\cite{p12-1}.
It is a rare SM process, and ATLAS measured its branching ratio with large statistics collected in 
2011 and 2012~\cite{p12-2}.  A clear signal is seen (see Fig.~\ref{Z4l}) and the measured branching
ratio of $(3.20 \pm 0.25 \stat \pm 0.13 \syst)\times 10^{-6}$ agrees with the SM prediction of $3.33 \times 10^{-6}$.
An earlier measurement by CMS from 7~\TeV~data is found here~\cite{p12-3}.
        \begin{figure}[h]
		         	\begin{minipage}[t]{0.49\linewidth}
\includegraphics[width=\linewidth]{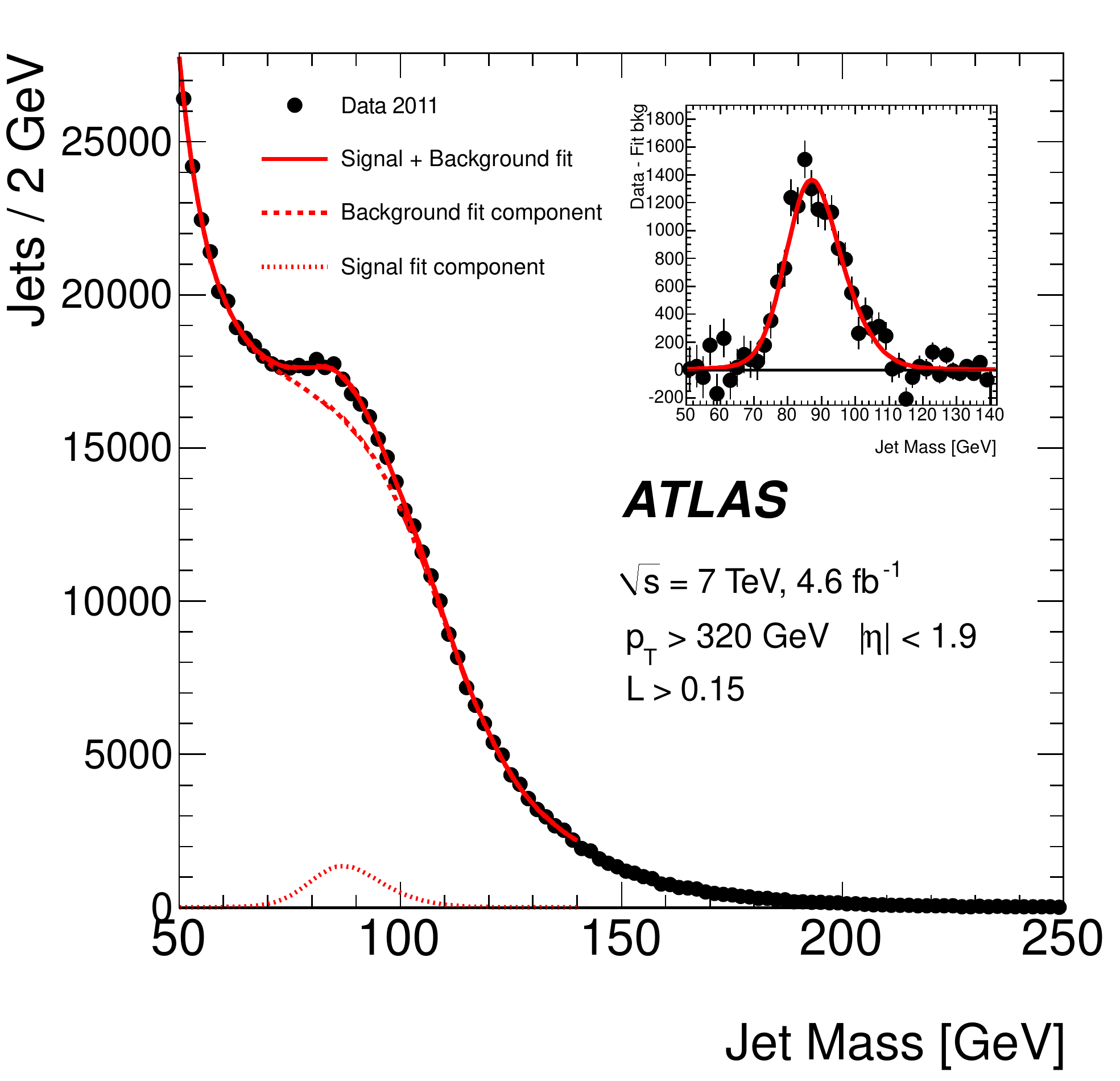}
\caption{Jet mass distribution for selected $W/Z$ jets overlaid with the fit result~\cite{p11-2}.}
\label{jetmass}
		\end{minipage}
				\hspace{0.01\linewidth}
		         	\begin{minipage}[t]{0.49\linewidth}
\includegraphics[width=\linewidth]{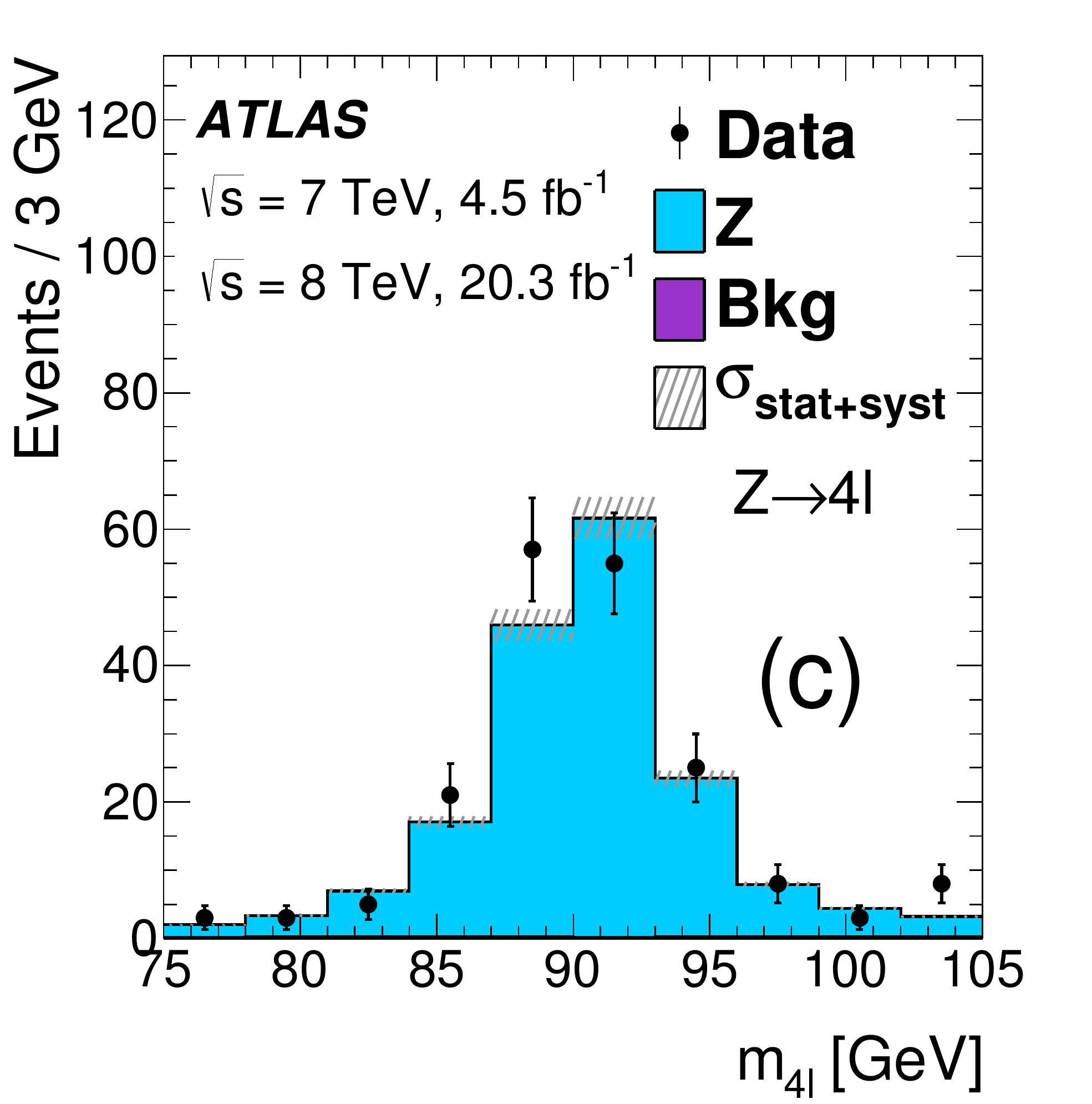}
\caption{Four-lepton invariant mass for the $Z \to 4\ell$ candidates~\cite{p12-2}.}
\label{Z4l}
		\end{minipage}
         \end{figure}	
         
\section{$W$+jets, $Z$+jets production}
Figure~\ref{Zjet} shows the distribution of inclusive number of jets produced in association with \Zboson
boson, measured by CMS from 8~\TeV~data~\cite{p13-1}.  The measurement extends to events with seven and
more jets.  The differential cross sections are obtained as functions of variables such as \pT and $\eta$ of the jets,
individually for jet multiplicities up to five.  Also double-differential cross sections are obtained in \pT and rapidity
of the jets~\cite{p13-2}.  Generally the NLO calculation using Sherpa2~\cite{Sherpa} describes the data better than the Leadng-Order (LO) prediction
from MadGraph~\cite{MadGraph}.  There is also an ATLAS measurement from 7~\TeV~\cite{p13-3} and a LHCb measurement with forward di-muons and jets~\cite{p13-4}.

Measurements of \Wboson production in association with jets are made by both ATLAS~\cite{p14-1} and
CMS~\cite{p14-2} with 7~\TeV~data.  NLO BlackHat~\cite{BlackHat}+Sherpa~\cite{Sherpa} does a good job in describing most
distributions, but it is a fixed-order calculation and thus higher jet multiplicities are missing.
Discrepancies with data are seen in distributions that are sensitive to this feature, for example
in the \ET sum of all jets.  This is improved in Sherpa2, as confirmed in the $Z$+jets analysis above.
Discrepancies are also observed in variables such as di-jet mass and azimuthal difference between the
jet and the muon from \Wboson.  The measurements are thus crucial in tuning the state-of-the-art
pQCD calculations and MC generators.

Since $W$+jets and $Z$+jets cross sections are measured, one can take a ratio
$R_{\rm jets} = \sigma_{\Wboson+\rm jets}/\sigma_{\Zboson+\rm jets}$ for the same jet kinematics.
Some experimental uncertainties and effects such as hadronization are largely reduced by
taking a ratio, so that the precisions are improved.
Figure~\ref{Rjet} shows $R_{\rm jets}$ as a function of the inclusive number of jets~\cite{p16}.
The first bin corresponds to the inclusive cross section without requiring jets (see Section~2)
and the ratio is slightly lower when requiring jets.
Generally BlackHat+Sherpa reproduces the data well, including the di-jet mass distribution
(in contrast to the individual measurements mentioned in the previous paragraph).
        \begin{figure}[h]
		         	\begin{minipage}[t]{0.49\linewidth}
\includegraphics[width=\linewidth]{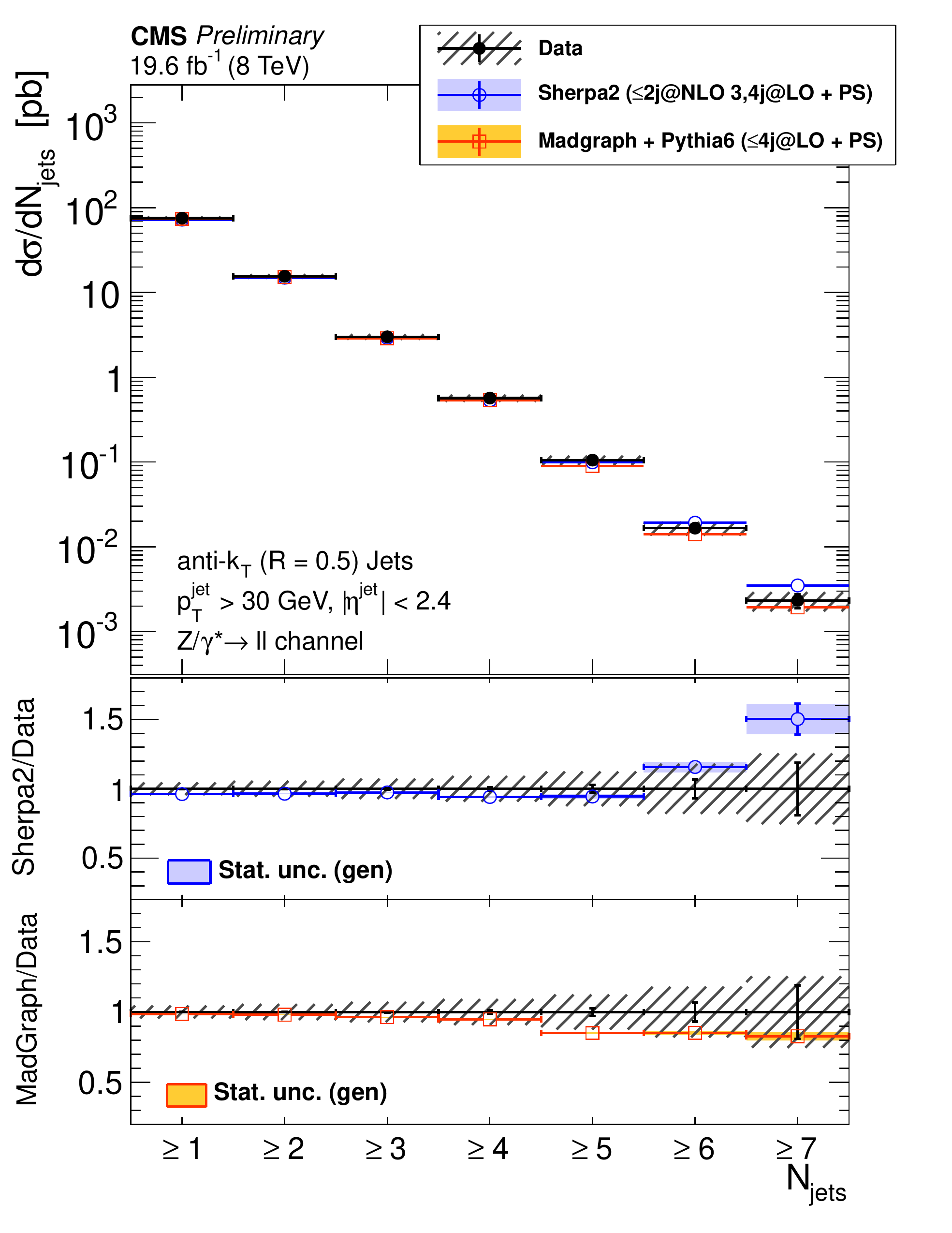}
\caption{$Z$+jets cross section as a function of inclusive jet multiplicity, compared with Sherpa2
and MadGraph predictions~\cite{p13-1}.}
\label{Zjet}
		\end{minipage}
				\hspace{0.01\linewidth}
		         	\begin{minipage}[t]{0.49\linewidth}
\includegraphics[width=\linewidth]{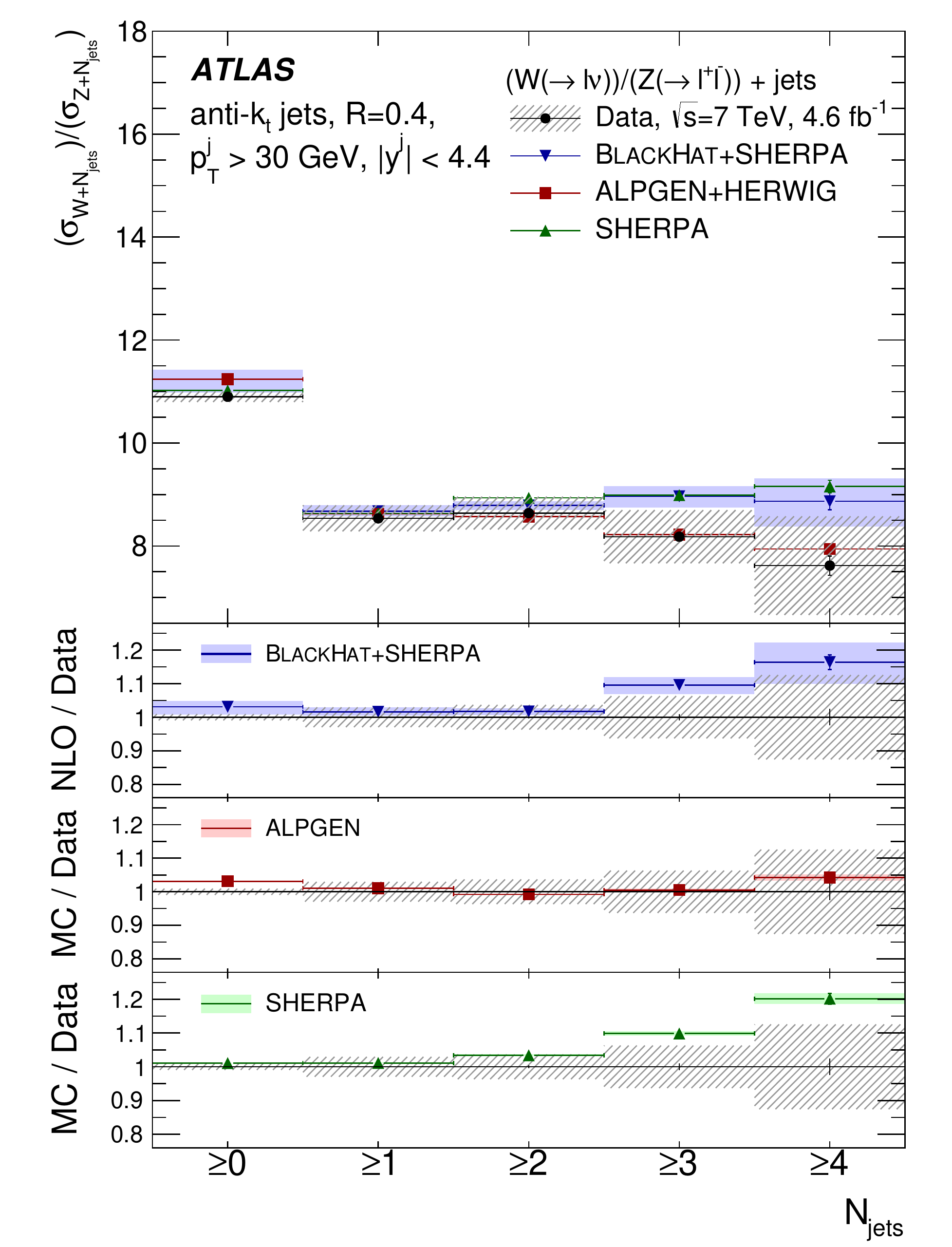}
\caption{Ratio of $W$+jets/$Z$+jets cross sections as a function of inclusive jet multiplicity~\cite{p16}.}
\label{Rjet}
		\end{minipage}
         \end{figure}	

\section{\Wboson/\Zboson+heavy flavor production}
The production of heavy flavor jets in association with \Wboson/\Zboson has a larger theoretical
uncertainty than light quark jets, thus its measurement is important.
The processes $Z(W)$+\bbbar~constitute also (irreducible) background to the
search for $ZH(WH)$ production followed by a $H \to \bbbar$ decay.
For the bottom-production calculation, there are two schemes, namely four flavor-number scheme (4FNS)
and five flavor-number scheme (5FNS), which treat $udsc+g$ and $udscb+g$ as the intrinsic parton
component of the proton, respectively.
The measured \Zboson~with one or more $b$-jets production cross section from ATLAS~\cite{p17-1}
seems to agree with 5FNS predictions better than 4FNS, while in the case of \Zboson with two or
more $b$-jets, the precisions are not good enough to draw a conclusion.
The MCFM~\cite{MCFM} fixed-order NLO calculation fails to describe the azimuthal difference between \Zboson
and $b$, while the prediction by aMC@NLO~\cite{aMCNLO} describes the data well, especially in 5FNS. (Fig.~\ref{Z+b}).
Also CMS measured $Z+b$ cross sections~\cite{p17-2}.

In the $W+\bbbar$ measurement by CMS~\cite{p18-1}, the largest background comes from the
\ttbar~production.  By requiring two well separated $b$-jets, the theoretical uncertainty is reduced.
The measured cross section agrees very well with the MCFM prediction.  ATLAS also made
a measurement of $W+b$ cross sections~\cite{p18-2}.

LHCb made a search for $Z$ and $D$-meson production, in which both particles are
reconstructed exclusively in $Z \to \mumu$ and $D^0 \to \Kminus \piplus$, $D^+ \to \Kminus \piplus \piplus$
(+charge conjugate) modes.  Signals of handful of events are observed as invariant mass peaks, and the obtained
cross sections are compared with the predictions from Single Parton Scattering and Double Parton Scattering
calculations~\cite{p19}.
        \begin{figure}[h]
		         	\begin{minipage}[t]{0.49\linewidth}
\includegraphics[width=\linewidth]{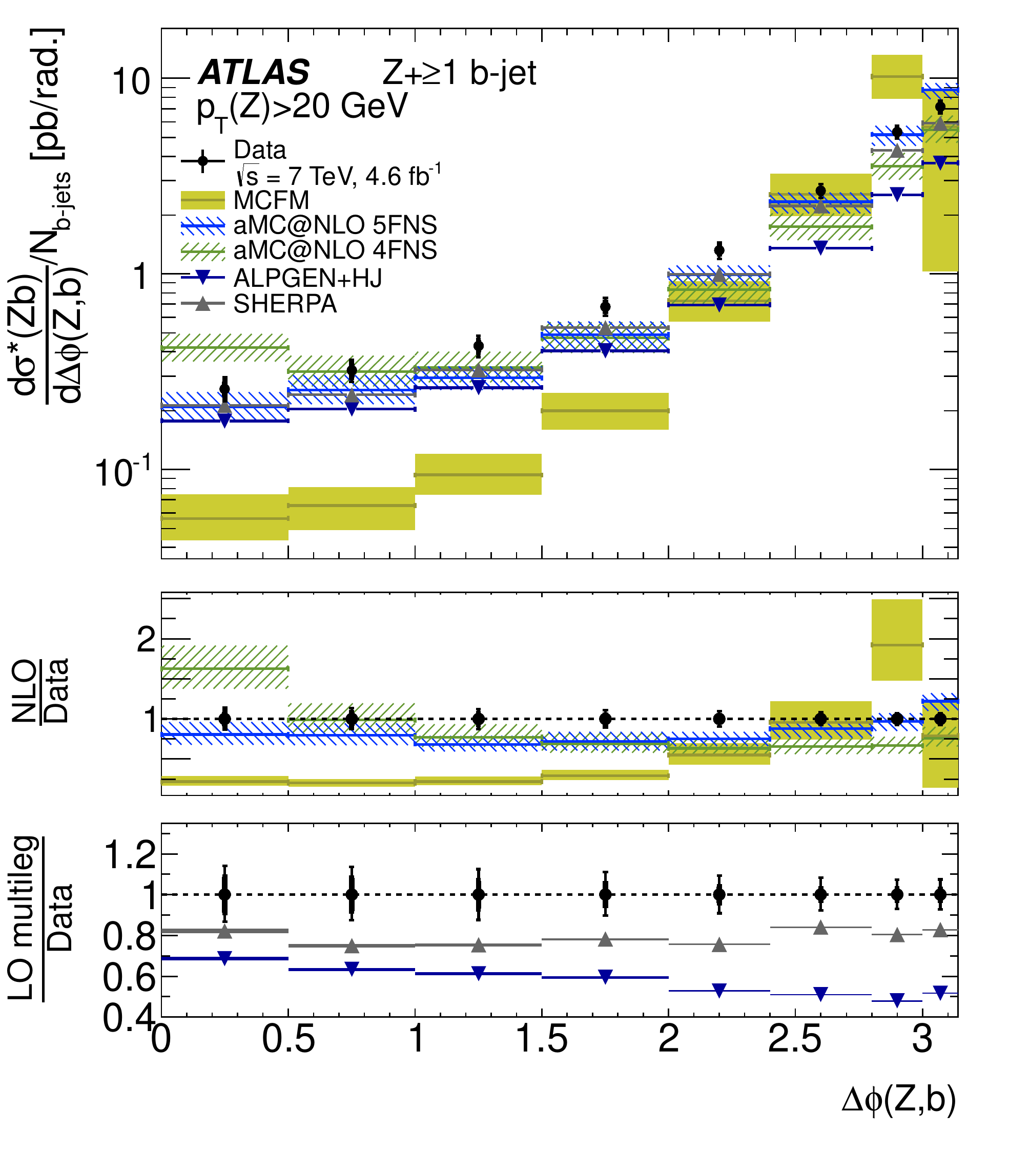}
\caption{Inclusive $b$-jet cross section as a function of azimuthal difference between $Z$ and $b$~\cite{p17-1}.}
\label{Z+b}
		\end{minipage}
				\hspace{0.01\linewidth}
		         	\begin{minipage}[t]{0.49\linewidth}
\includegraphics[width=\linewidth]{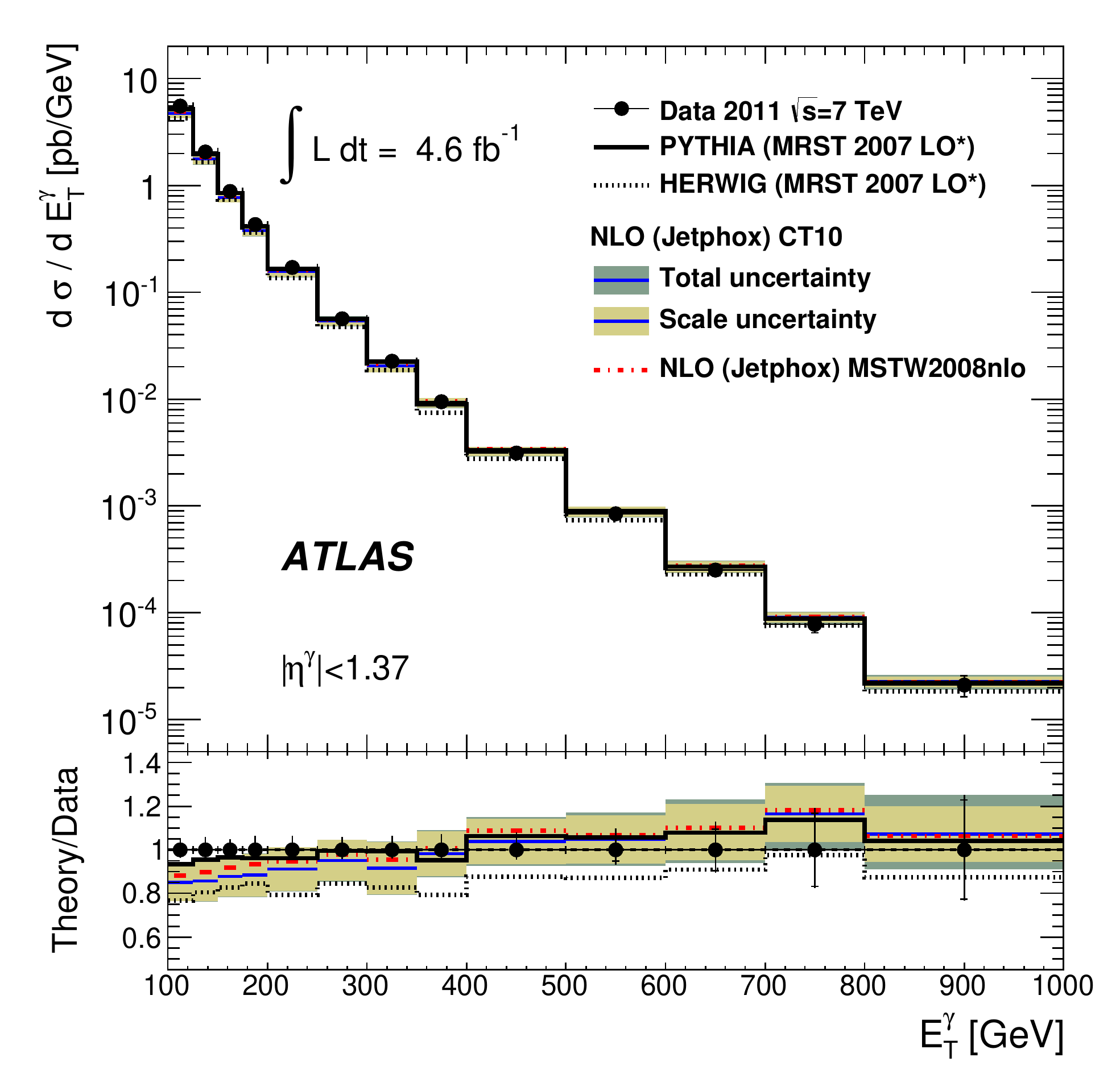}
\caption{Inclusive prompt photon cross section as a function of photon \ET in the barrel region~\cite{p20}.}
\label{Prompt}
		\end{minipage}
         \end{figure}	

\section{Prompt $\gamma$ and $\gamma\gamma$ pair production, $\gamma$+jets}
The lowest-order processes of prompt photon production are
$qg \to q\gamma, \qqbar \to g\gamma$.
The measurements are sensitive to the gluon PDF.
To extract the signal from misidentified hadrons and non-prompt photons,
the isolation energy around the photon candidate is used.
In the ATLAS measurement~\cite{p20}, the maximum photon \ET reaches about 1~\TeV~(see Fig.~\ref{Prompt}), and the distributions are
compared with NLO calculations and LO MC generators.
Both describe the data distributions well, although the predictions are lower than the data by about 20\%
in magnitude. 

The pair production of photons occur by $\qqbar \to \gamma\gamma$ in LO, and contributes
to the background in the $H \to \gamma\gamma$ measurement.
Both CMS~\cite{p21-1} and ATLAS~\cite{p21-2} measured various distributions, with \ET
thresholds for the two photons at (40, 25)\GeV~and (25, 22)\GeV, respectively.
Among the predictions compared to data, Sherpa and 2$\gamma$NNLO~\cite{2gNNLO} describe the data pretty well.
         \begin{figure}[h]
			\vspace{-3.5mm}
		         	\begin{minipage}[t]{0.49\linewidth}
			\vspace{-7.3cm}
			\includegraphics[width=\linewidth,angle=0]{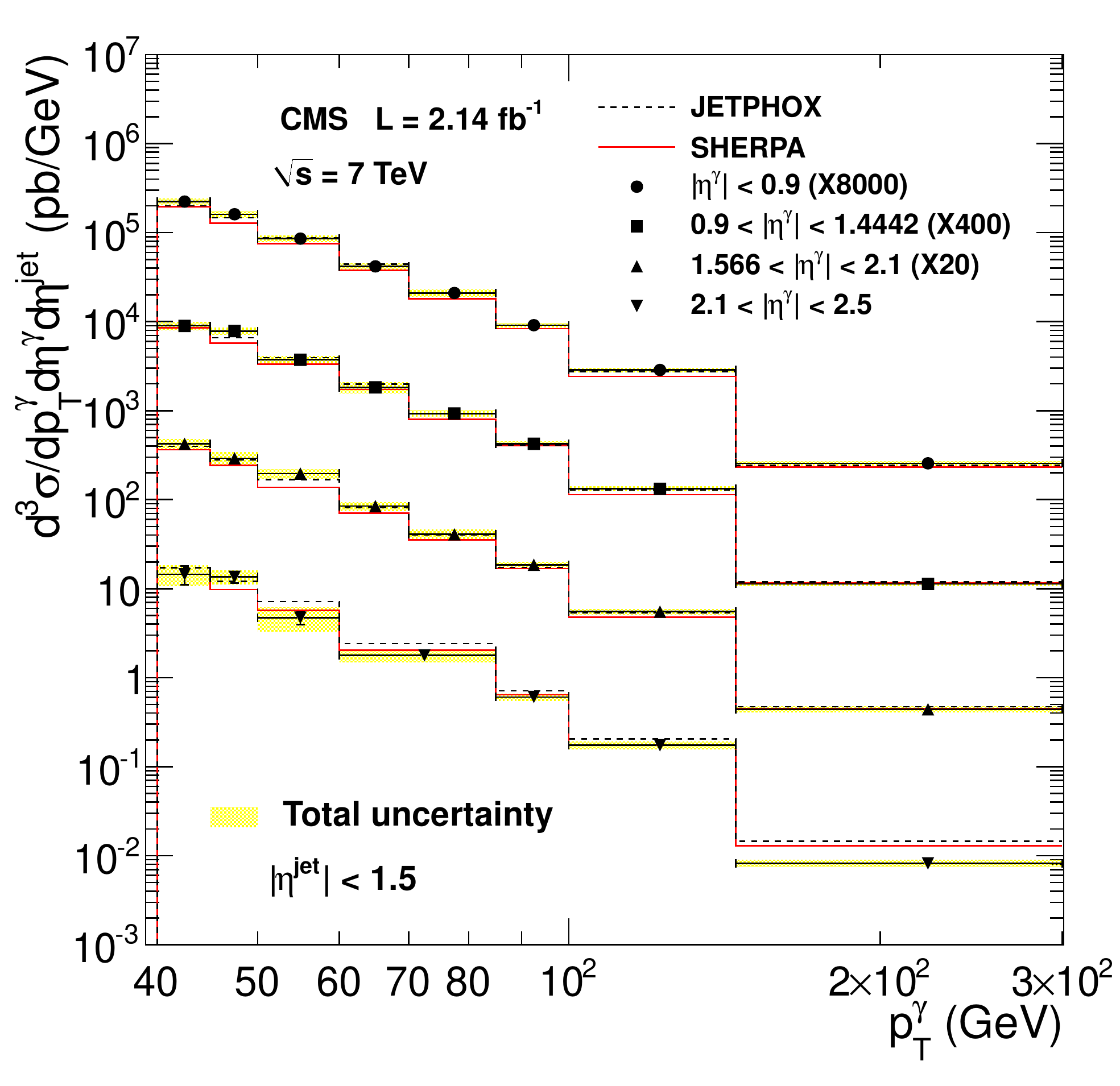}
\caption{Photon+jets differential cross sections as functions of \pT and $\eta$ of $\gamma$,
for $|\eta^{\rm jet}| < 1.5$~\cite{p22}.}
\label{Gjet}
		\end{minipage}
				\hspace{0.01\linewidth}
		         	\begin{minipage}[t]{0.49\linewidth}
\includegraphics[width=\linewidth]{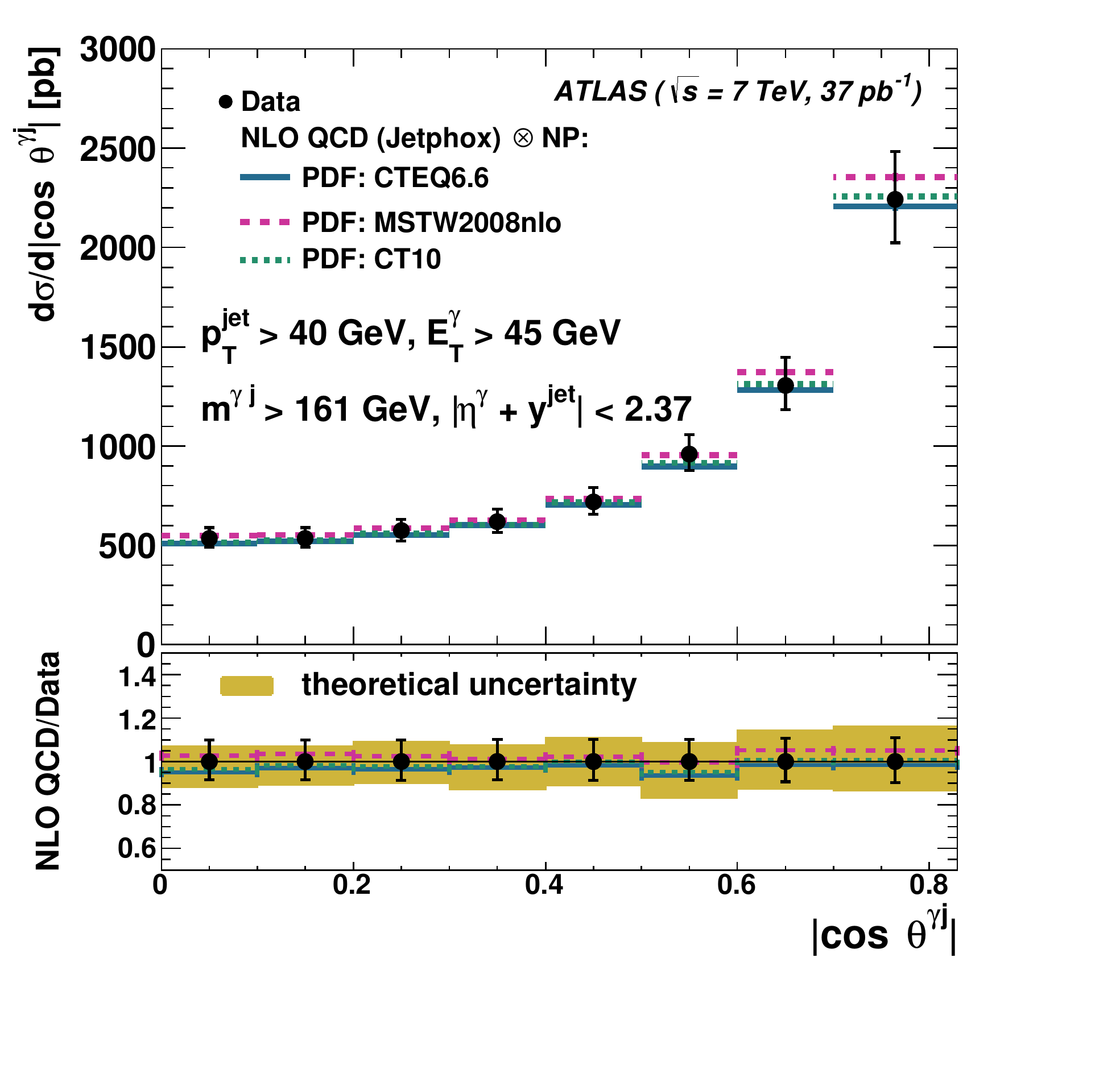}
\caption{Isolated-photon cross section as a function of $|\cos\theta^{\gamma \rm j}|$ (see text for definition)~\cite{p23}.}
\label{cosT}
		\end{minipage}
         \end{figure}	

CMS measured the $\gamma$+jets triple-differential cross sections in \pT, $\eta$ of the photon and $\eta$
of the jet~\cite{p22}.  Predictions from Jetphox~\cite{Jetphox} and Sherpa agree with the data, except for the highest
photon \pT and largest photon $\eta$ (see Fig.~\ref{Gjet}).  
The ATLAS analysis~\cite{p23} investigated the dynamics in $\gamma$+jets production.
In the $qg \to q\gamma$ process, a spin-1/2 $t$-channel quark propagator is exchanged,
while in the di-jet production, the propagator is dominantly spin-1 gluon.
This is illustrated in the distribution of $|\cos\theta^{\gamma \rm j}|$, which is the scattering angle
in the two-parton center-of-mass system, as shown in Fig.~\ref{cosT}.
Concerning the azimuthal separation between $\gamma$ and jet,
the NLO calculation (Jetphox) has difficulty in predicting events with small separation,
since the $\gamma$ and jet cannot be in the same hemisphere in the calculation by construction.
For this distribution LO PYTHIA with parton shower describes the data well.

Finally, the $\gamma$+jets and $Z$+jets measurements are combined together.
In the CMS analysis on rapidity distribution~\cite{p24}, the (absolute) average of
vector-boson rapidity and jet rapidity ($y_{\rm sum}$) and the (absolute) difference
of the two ($y_{\rm dif}$) are investigated.
$y_{\rm sum}$ reflects the total boost of the system, and is sensitive to the PDF, while
$y_{\rm dif}$ reflects the scattering dynamics, for which various predictions differ
at large $y_{\rm dif}$ values.
For both \Zboson and $\gamma$, the NLO predictions (by MCFM and Owens~\cite{Owens}, respectively)
agree well with the data at large $y_{\rm dif}$ value, as shown in Fig.~\ref{ydif}.
       \begin{figure}[h]
			\vspace{-4.5mm}
		         	\begin{minipage}[t]{0.49\linewidth}
\includegraphics[width=\linewidth]{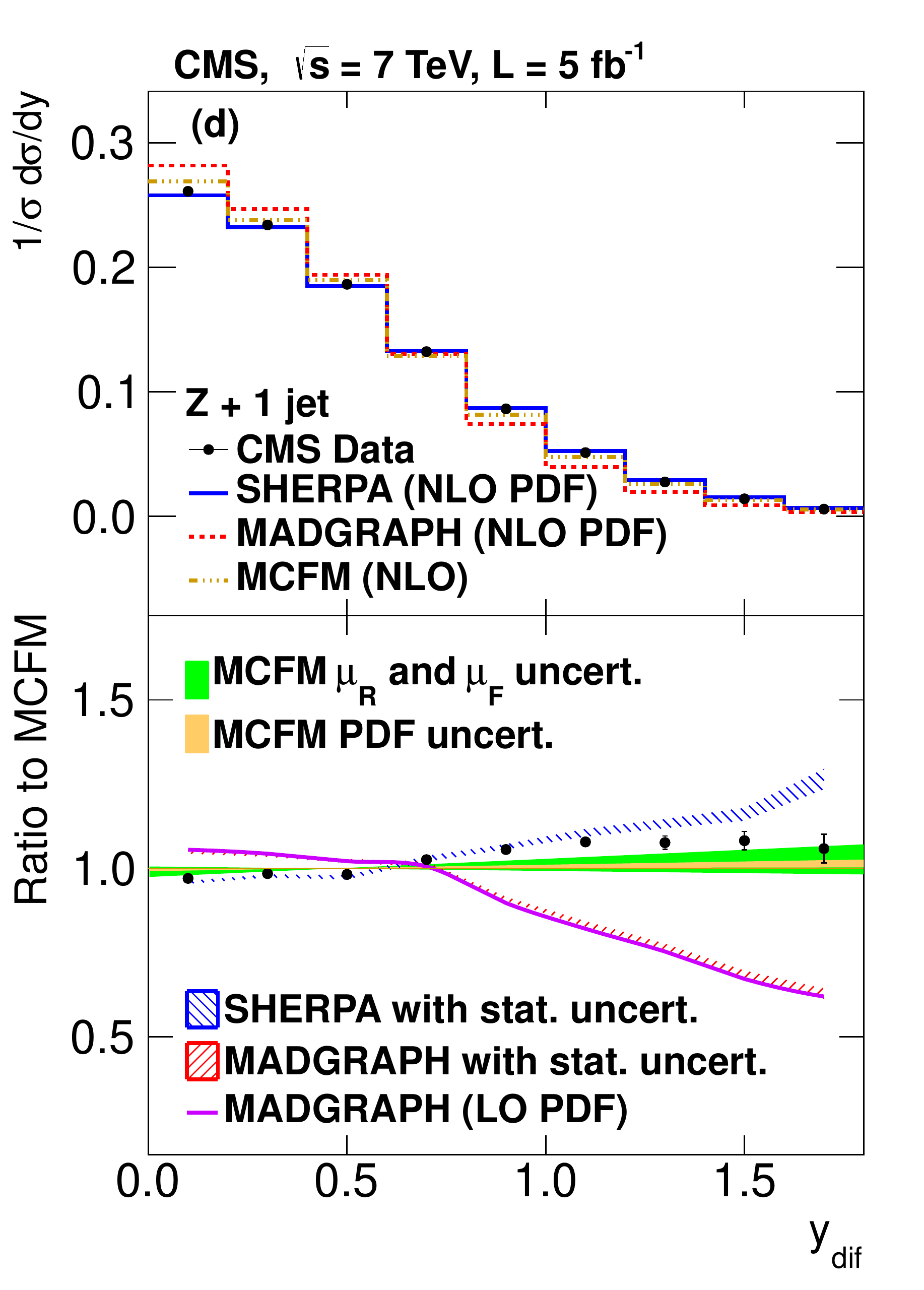}
\caption{Normalized distribution of $y_{\rm dif} \equiv |y_Z - y_{\rm jet}|/2$ in $Z$+1 jet events~\cite{p24}.}
\label{ydif}
		\end{minipage}
				\hspace{0.01\linewidth}
		         	\begin{minipage}[t]{0.48\linewidth}
			\vspace{-5.8cm}
			\hspace{-1cm}
\includegraphics[width=\linewidth,angle=0]{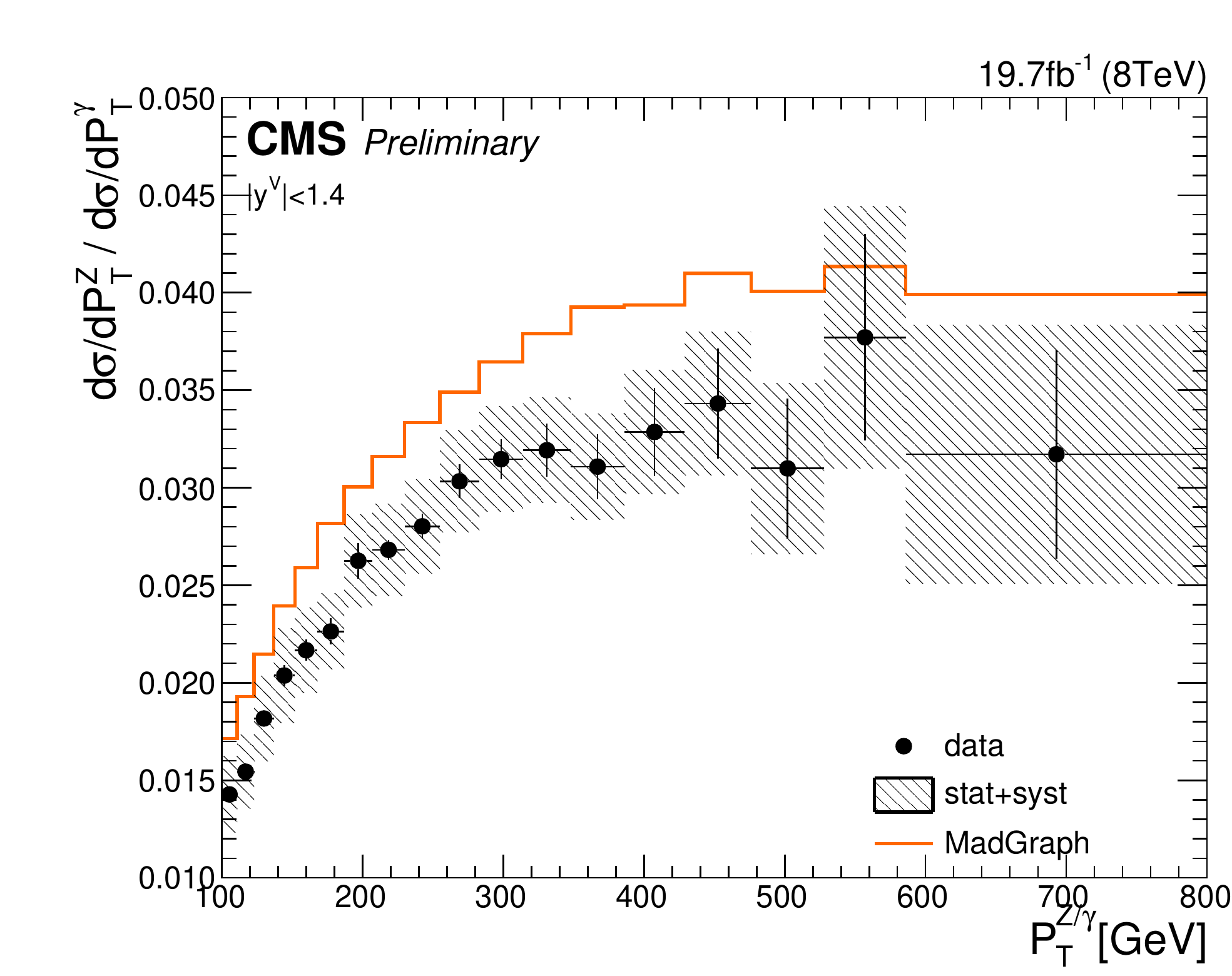}
\caption{Differential cross-section ratio of leptonic \Zboson~over $\gamma$ as a function of the
vector boson \pT~\cite{p25}.}
\label{ZGratio}
		\end{minipage}
         \end{figure}	

CMS also compared the \Zboson/$\gamma^*$+jets and $\gamma$+jets cross sections
at the same vector boson kinematics~\cite{p25}.  The ratio of \Zboson~to $\gamma$ reaches a plateau,
at about 0.03,  for \pT of the vector boson 
$\gtrsim$ 300~\GeV, as shown in
Fig.~\ref{ZGratio}.
MadGraph reproduces its shape.
This fact is useful in the new physics search with \MET and jets, in which
the irreducible background comes from (\Zboson $\to$ \nnbar)+jets process.
It validates the estimation of such background from the $\gamma$+jets event kinematics.

\section{Conclusion}
A wealth of results has been obtained in \Wboson/\Zboson/$\gamma$ production at the LHC
from Run 1 data.
The SM (electroweak+QCD) has been tested at the highest energy for extreme kinematics
(up to $\approx 1~\TeV$) and topologies (with many jets, heavy flavors, etc.).
These are crucial inputs for developments of theoretical predictions (NLO, NNLO, etc.) and
MC generators, which show the readiness for new physics searches in Run~2, starting in 2015.

%



\end{document}